\documentclass[
reprint,
superscriptaddress,
amsmath,amssymb,
showpacs,
prb,
aps,
]{revtex4-1}

\usepackage{graphicx}
\usepackage{dcolumn}
\usepackage{bm}
\usepackage{color} 
\usepackage{ulem} 
\usepackage{notoccite}
\usepackage{epstopdf}
\usepackage{hyperref}
\hypersetup{%
  colorlinks=true,
  linkcolor=blue,
  citecolor=blue,
  urlcolor=blue,
}

\newcommand*{\citen}[1]{%
  \begingroup
    \romannumeral-`\x 
    \setcitestyle{numbers}%
    \cite{#1}%
  \endgroup   
}

\begin{document}

\title{Phonon Density of States and Anharmonicity of UO$_\textnormal{2}$}

\author{Judy W.L. Pang}
 \email{pangj@ornl.gov}
 \affiliation{Oak Ridge National Laboratory, Materials Science and Technology Division,  Oak Ridge, Tennessee 37831, USA}
\author{Aleksandr Chernatynskiy}
\affiliation{Department of Materials Science and Engineering,~University of  Florida, Gainesville, Florida 32611, USA}
\author{Bennett C. Larson}
 \affiliation{Oak Ridge National Laboratory, Materials Science and Technology Division,  Oak Ridge, Tennessee 37831, USA}
\author{\mbox{William J.L. Buyers}}
\affiliation{Chalk River Laboratories, National Research Council, Chalk River, ON, K0J 1J0 Canada}
\author{Douglas L. Abernathy}
 \affiliation{Oak Ridge National Laboratory, Quantum Condensed Matter Division, Oak Ridge, Tennessee 37831, USA}
\author{Kenneth J. McClellan}
\affiliation{Los Alamos National Laboratory, Materials Science and Technology Division, Los Alamos, New Mexico 87545, USA}
\author{Simon R. Phillpot}
\affiliation{Department of Materials Science and Engineering,~University of  Florida, Gainesville, Florida 32611, USA}

\date{\today}

\begin{abstract}
Phonon density of states (PDOS) measurements have been performed on polycrystalline UO$_\textnormal{2}$  at 295 and 1200 K using time-of-flight inelastic neutron scattering to investigate the impact of anharmonicity on the vibrational spectra and to benchmark \textit{ab initio} PDOS simulations performed on this strongly correlated Mott-insulator. Time-of-flight PDOS measurements include anharmonic linewidth broadening inherently and the factor of $\sim$ 7 enhancement of the oxygen spectrum relative to the uranium component by the neutron weighting increases sensitivity to the oxygen-dominated optical phonon modes. The first-principles simulations of quasi-harmonic PDOS spectra were neutron-weighted and anharmonicity was introduced in an approximate way by convolution with wavevector-weighted averages over our previously measured phonon linewidths for UO$_\textnormal{2}$ that are provided in numerical form. Comparisons between the PDOS measurements and the simulations show reasonable agreement overall, but they also reveal important areas of disagreement for both high and low temperatures. The discrepancies stem largely from an $\sim$ 10 meV compression in the overall bandwidth (energy range) of the oxygen-dominated optical phonons in the simulations. A similar linewidth-convoluted comparison performed with the PDOS spectrum of Dolling \textit{et al.} obtained by shell-model fitting to their historical phonon dispersion measurements shows excellent agreement with the time-of-flight PDOS measurements reported here. In contrast, we show by comparisons of spectra in linewidth-convoluted form that recent first-principles simulations for UO$_\textnormal{2}$ fail to account for the PDOS spectrum determined from the measurements of Dolling \textit{et al.} These results demonstrate PDOS measurements to be stringent tests for \textit{ab inito} simulations of phonon physics in UO$_\textnormal{2}$  and they indicate further the need for advances in theory to address lattice dynamics of UO$_\textnormal{2}$.  
\end{abstract}
\maketitle

\section{INTRODUCTION}
The physical properties of UO$_\textnormal{2}$ are of strong scientific interest \textemdash\ fundamentally as a highly correlated electronic system and technologically as the most widely used nuclear fuel. As a Mott-insulator, the technologically important thermal transport of UO$_\textnormal{2}$ is controlled by phonon kinetics and anharmonicity, which are in turn sensitive functions of the strongly correlated electronic structure of UO$_\textnormal{2}$.~\cite{yin2008,pang2013} While recent first-principles simulations for UO$_\textnormal{2}$~\cite{sanati2011,yun2012,wang_b2013} have reported good agreement with experimental elastic constants, bulk moduli, and lattice heat capacity measurements, a reduced level of agreement was found for phonon dispersion and phonon density of states (PDOS) simulations compared to the experimental inelastic neutron scattering (INS) results of Dolling \textit{et al.}~\cite{dolling1965}  Moreover, Pang \textit{et al.}~\cite{pang2013} reported results similar to other \textit{ab initio} phonon dispersion simulations,~\cite{pang2013,sanati2011,yun2012,wang_b2013} but differences of more than a factor of two between \textit{ab initio} simulations and INS measurements of phonon linewidths (i.e. inverse lifetimes) for UO$_\textnormal{2}$,~\cite{pang2013} especially at high temperature (1200 K). 

Since 3$^\textnormal{rd}$ order interatomic forces are required for simulations of phonon linewidths and thermal transport,~\cite{pang2013} it might be anticipated that the large discrepancies in the linewidths are a result of the difficulty in calculating 3$^\textnormal{rd}$ order derivatives for the strongly correlated 5\textit{f} electronic structure of UO$_\textnormal{2}$. Counter-intuitively, however, the major source of the discrepancy in the phonon linewidth simulations was identified~\cite{pang2013} to lie within the phonon energies and dispersion. This suggests that presently available \textit{ab initio} simulations~\cite{sanati2011,yun2012,wang_b2013} of the 2$^\textnormal{nd}$ order interatomic forces are not sufficiently accurate for handling  anharmonicity and phonon linewidths, and hence, for simulations of thermal transport in UO$_\textnormal{2}$. Considering the large anharmonicity of UO$_\textnormal{2}$ and the limitations of the quasi-harmonic approximation to account for temperature effects as used in Ref.~\citen{pang2013}, other methods such as \textit{ab initio} molecular dynamics \cite{thomas2010,lan2012} that include anharmonicity rigorously should be considered as well.

The strong sensitivity of phonon linewidth simulations to the 2$^\textnormal{nd}$ order force derivatives and phonon dispersion has been discussed in the context of \textit{ab initio} simulations of phonon lifetimes and thermal transport in non-highly correlated systems,~\cite{broido2007,tang2010a,tian2012,ward2010} in which it has been shown that highly accurate dispersion simulations can be performed. Since phonon linewidths have a collective inverse 4$^\textnormal{th}$ power dependence on phonon energies within 3-phonon scattering,~\cite{ward2010,maradudin1962} it is important to make make quantitative experimental tests of the accuracy of phonon energies in the simulations for UO$_\textnormal{2}.$~\cite{pang2013,sanati2011,yun2012,wang_b2013} These first-principles PDOS simulations have been compared with the shell model derived PDOS specturm of Dolling based on symmetry direction dispersion measurements for UO$_\textnormal{2}.$~\cite{dolling1965} However, until now, neutron time-of-flight measurements of PDOS have not been available to provide full Brillouin zone tests of phonon zone boundary energies and phonon energy gradients,~\cite{thomas2010} as well as the impact of temperature and anharmonicity on the spectrum. In fact, the only measured PDOS spectrum reported for an actinide oxide was measured by inelastic x-ray scattering at room temperature in polycrystalline PuO$_\textnormal{2}$~\cite{manley2012} because of the strong thermal neutron absorption of Pu.
 
To the extent that phonon dispersion accuracy comparable to that achieved for weakly correlated materials~\cite{broido2007,tang2010a,tian2012,ward2010} is required for phonon transport simulations, further extension of the DFT approach such as through hybrid functional density functional theory (HF-DFT),~\cite{jollet2009,kudin2002,wen2012} or through higher-level first-principles theories such as dynamical mean field theory (DMFT)~\cite{yin2008,yin2011,dorado2013} may be needed for UO$_\textnormal{2}$.
Unfortunately, calculations of 3$^\textnormal{rd}$ order interatomic forces necessary for phonon linewidth simulations are likely to be computationally prohibitive for these techniques using the finite difference approach and will be technically very challenging for faithful implementation of the analytical derivatives of the total energy. Nevertheless, we emphasize that phonon dispersion has been found~\cite{pang2013} to have a strong influence on the determination of phonon lifetimes in UO$_\textnormal{2}$ so accuracy of phonon energies is critical. We note also that new numerical approaches are under development for finite temperature lattice dynamics from first principles.~\cite{chen2013}
 
Here we report time-of-flight INS PDOS measurements on UO$_\textnormal{2}$ at 295 and 1200 K. We also report detailed comparisons of these measurements with \textit{\textit{ab initio}} PDOS simulations by introducing anharmonic broadening into the simulations through energy averages over our previously reported comprehensive set of experimental phonon linewidths for UO$_\textnormal{2}$.~\cite{pang2013} We further compare our phonon-linewidth-convoluted PDOS simulations with phonon linewidth broadened PDOS simulations in the literature~\cite{sanati2011,yun2012,wang_b2013}  and with the shell model derived PDOS spectrum of Dolling \textit{et al.}~\cite{dolling1965} for UO$_\textnormal{2}$. These comparisons show excellent agreement between our time-of flight measurements and the phonon-broadened shell model derived PDOS spectrum of Dolling, but demonstrate significant and systematic discrepancies between experimental PDOS measurements and presently available \textit{ab initio} simulations.

\section{EXPERIMENT}
Neutron-weighted phonon density of states measurements were determined  for UO$_\textnormal{2}$ at 295 K and 1200 K through INS measurements on the ARCS time-of-fight spectrometer at the Spallation Neutron Source (SNS) at Oak Ridge National Laboratory.~\cite{abernathy2012} The samples of depleted UO$_\textnormal{2}$ polycrystalline material were sintered to \mbox{92 \%} of theoretical density at 1650$^\circ$C under Ar followed by conditioning at temperature and cool down under Ar-6\% H$_{2}$ in order to achieve a nominal oxygen stoichiometry of 2.00. The measurements were performed on $\sim$ 5 mm diameter cylindrical samples with a total mass of 9.09 g encapsulated in a 5.5 mm diameter vanadium can mounted in a high temperature vacuum furnace. Measurements on the ARCS spectrometer with incident neutron energies, E$_{i}$, of 30, 60, and 120 meV provided energy resolutions (FWHM) of $\sim$ 0.9, 1.8, and 3.6 meV, respectively, as described in detail elsewhere.~\cite{abernathy2012} The backgrounds introduced by the sample containment can and the furnace were removed using measurements on a duplicate (but empty) vanadium can. The detector efficiency of the spectrometer was calibrated by measurements on reference vanadium powder. 

The corrected spectra were normalized by the incident proton current of 4 Coulomb$/$hr and binned to obtain the scattered neutron counts $I(E,\phi)$ as a function of energy loss \textit{E} and scattering angle $\phi$. To obtain spectra within the incoherent scattering approximation, the scattered intensity spectra were converted to the scattering function $S(E,Q)$ and integrated over momentum transfers $Q$ ranging from 3 to 7 \r{A}$^{-1}$, corresponding to about four Brillouin zones of UO$_\textnormal{2}$ to obtain $S(E)$, which was then converted to the neutron-weighted generalized PDOS $g^{NW}(E)$.~\cite{fultz2010} Because of the form factor, the impact of the magnetic moments of the U$^{4+}$ ions is not significant for  $Q$ above 3 \r{A}$^{-1}$. Corrections for multi-phonon and multiple scatterings effects~\cite{delaire2009,dhital2012} were made using the iterative method~\cite{kresch2008} of Sears \textit{et al.}~\cite{sears1995}  In order to take advantage of the higher energy resolution of the energy loss spectra with lower incident neutron energies, the phonon energy spectrum was covered by concatenating measurements for \textit{E} below 24 meV from the E$_{i}$ = 30 meV spectrum, the density of states between \textit{E} = 24  \textendash\ 45 meV was obtained from the E$_{i}$ = 60 meV spectrum, and the E$_{i}$ = 120 meV spectrum was used to determine the intensities for \textit{E} above 45 meV.  Using the method as described in Ref.~\citen{dhital2012}, the full range of the spectrum was determined by normalizing the total intensity of the E$_{i}$ = 30 meV spectrum to the percentage of the density of states below 24 meV of the 60 meV spectrum, followed by the normalization of the total weight of the subsequent spectrum  to the fraction of the density of states below 45 meV of the E$_{i}$ = 120 meV spectrum.

For inelastic scattering averaged over this wide range (3 \textendash\ 7 \r{A}$^{-1}$) of $|Q|$,~\cite{fultz2010,loong1991} the neutron-weighted PDOS spectrum for polycrystalline UO$_\textnormal{2}$ is given by:
\begin{equation}
	g^{NW} \cong \frac{\sigma_{U}}{M_{U}}g_{U}(E) + 2\frac{\sigma_{O}}{M_{O}}g_{O}(E),
	\label{eq:PDOS_NW}
\end{equation}

where $g_U(E)$ and $g_O(E)$ are the partial density of states for uranium and oxygen in UO$_\textnormal{2}$, respectively, and $M_{i}$ and $\sigma_{i}$ are the atomic masses and neutron scattering cross-sections~\cite{NIST} for the $i$ = U and O atoms. Since $\frac{\sigma_{O}}{M_{O}}$ is about 7 times $\frac{\sigma_{U}}{M_{U}}$, the neutron-weighted PDOS spectra for UO$_\textnormal{2}$ is inherently weighted heavily toward oxygen atom scattering. We note that the simple form of Eq. (\ref{eq:PDOS_NW}) is possible because the effect on the measured PDOS spectrum of the thermal Debye-Waller factors $exp(-2W_{i})$ for $i$ = U, O is known to be negligible for low temperatures.~\cite{fultz2010} And we have found that even at 1200 K the overall impact of the Debye-Waller factors (using the measured Debye-Waller factors by Willis~\cite{willis1963}) for UO$_\textnormal{2}$ is under 2 \%, which is within the experimental uncertainty of the measured PDOS spectra. All the presented PDOS spectra are normalized to unity. 

\section{RESULTS}
\subsection{Time-of-Flight PDOS Measurements at 295 K and 1200 K}

\begin{figure}
\includegraphics[width=3.3in]{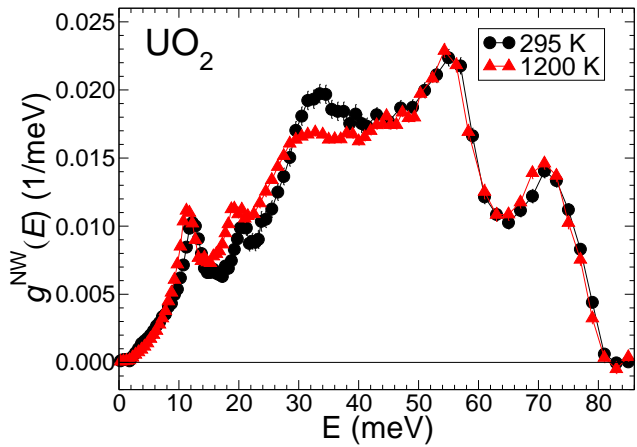}
\caption{\label{fig:PDOS} Neutron-weighted phonon density of states {\textit{g}}$^{\textnormal{NW}}$(\textit{E}) of UO$_\textnormal{2}$ measured at 295 K and 1200 K.}
\end{figure}

\begin{figure*}
\includegraphics[width=6.5in]{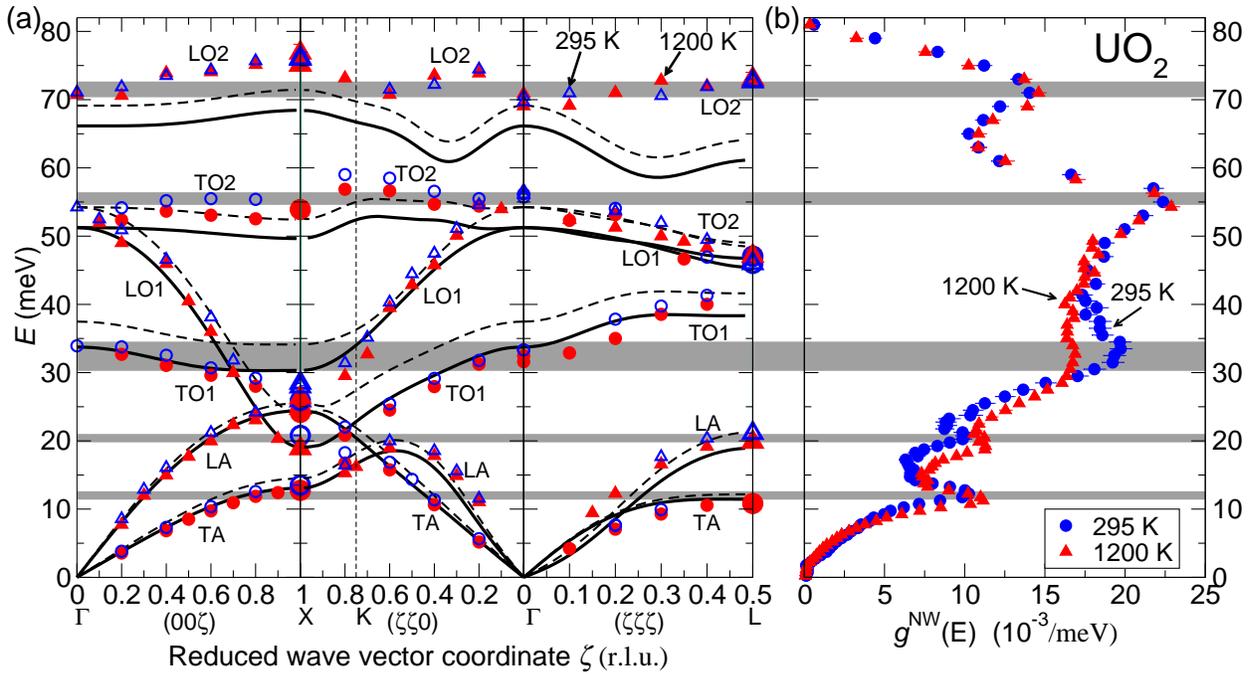}
\caption{\label{fig:Dispersion} (a) Phonon dispersion curves of UO$_\textnormal{2}$ measured at 295 K (blue open symbols) and at 1200 K (red solid symbols). Circles (triangles) represent the transverse (longitudinal) phonons. Measurements at the zone boundary points X and L are highlighted with larger symbols. Dashed and solid lines are the simulations of phonon dispersion at 295 and 1200 K respectively, using GGA + \textit{U} approximations. (b) Corresponding neutron-weighted phonon density of states {\textit{g}}$^{\textnormal{NW}}$(\textit{E}) of UO$_\textnormal{2}$ measurements at 295 K and 1200 K.
}
\end{figure*}

Figure~\ref{fig:PDOS} shows experimental phonon density of states spectra for UO$_\textnormal{2}$ measured at temperatures of 295 K and 1200 K. The 295 K spectrum has resolved peaks at (nominally) 12, 21, 33, 56, and 72 meV and a cutoff at 81 meV while the corresponding peaks at 1200 K were found to be at 11, 19, 30, 55, and 70 meV, with a cutoff at 81 meV. As illustrated in Fig.~\ref{fig:Dispersion}(a), these peaks correspond to zone boundary phonon energies identified in previously reported single crystal UO$_\textnormal{2}$ measurements at 295 and 1200 K;~\cite{pang2013} these data are plotted in Fig.~\ref{fig:Dispersion}(a) for completeness and for zone-boundary symmetry direction identification. The peak at 11 \textendash\ 12 meV corresponds to transverse acoustic (TA) zone-boundary phonons at the X and L points and the weaker (and broader) peak at $\sim$ 20 meV corresponds to the longitudinal acoustic (LA) phonon zone boundary energies at the K and L points. The broad PDOS peak for 295 K and the marginally resolved peak for 1200 K at 33 meV correspond roughly to the TO1 transverse optical and LO1 longitudinal optical zone boundary energies at the X and K points. The peak at 56 meV stems from the TO2 transverse optical zone boundary energies at the X and K points and the 72 meV peak corresponds to the slowly dispersing LO2 longitudinal optical phonon zone boundary energies at the X, K, and L points. The relatively small ($<$2 meV) softening observable in both the two low-energy PDOS peaks and the two high-energy peaks in Fig.~\ref{fig:PDOS} and Fig.~\ref{fig:Dispersion}(b), i.e. as the temperature was changed from 295 to 1200 K, is consistent with the similarly small changes observed in the high symmetry direction single crystal dispersion measurements in Fig.~\ref{fig:Dispersion}(a).

The most noticeable temperature-induced phonon lifetime change in the PDOS lies in the intermediate energy range of 25 \textendash\ 40 meV where the already broad, but well-resolved peak at 33 meV is broadened at 1200 K to an almost unresolved shoulder in the spectrum. This is in contrast to little or no observable change in the shape of the peaks at 56 and 72 meV where the ARCS spectrometer measurement resolution is the widest ($\sim$ 3 meV). 

\subsection{First-Principles PDOS Simulations}
\begin{figure}
\includegraphics[width=3.3in]{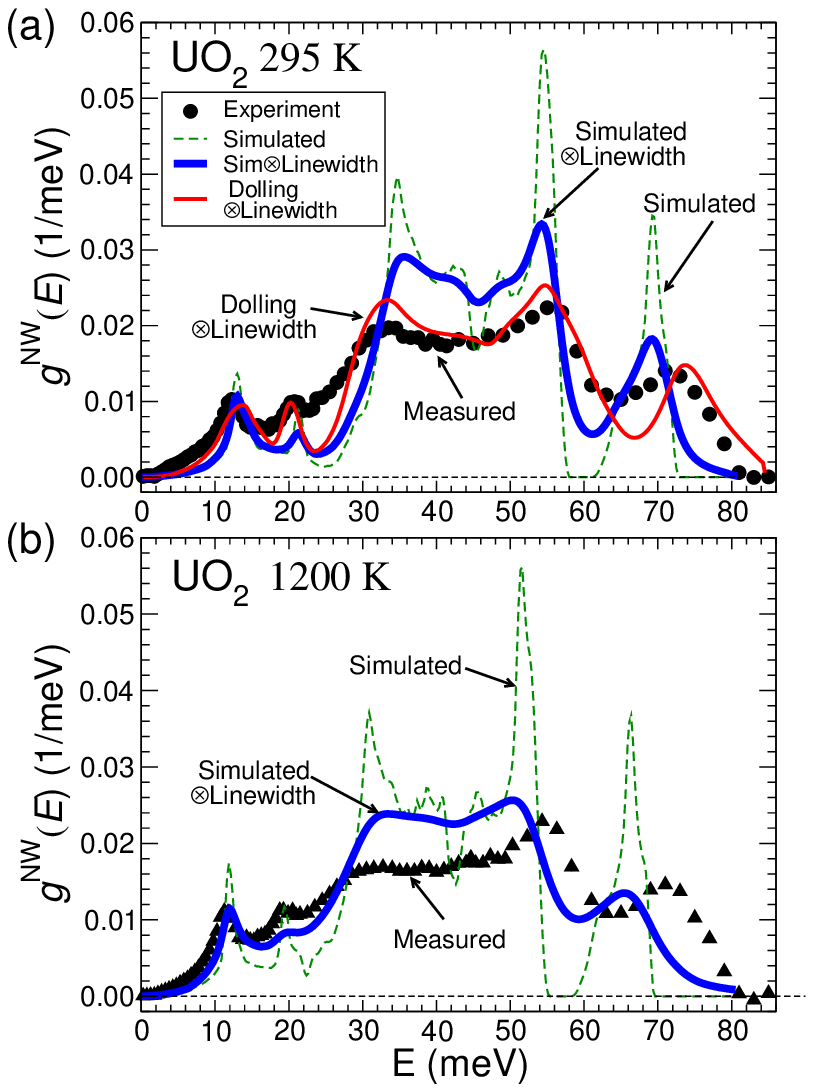}
\caption{\label{fig:PDOS_ExptDFT} Neutron-weighted phonon density of states {\textit{g}}$^{\textnormal{NW}}$(\textit{E}) of UO$_\textnormal{2}$ at (a) 295 K and (b) 1200 K by neutron scattering (symbols); quasi-harmonic simulations (thin green lines); and quasi-harmonic simulations convoluted with intrinsic phonon linewidths (thick blue lines). Instrumental resolution is included in all the simulations.}
\end{figure}

To test our fundamental understanding of UO$_\textnormal{2}$ using the above PDOS measurements, we show in Figs.~\ref{fig:PDOS_ExptDFT}a and~\ref{fig:PDOS_ExptDFT}b \textit{ab initio} PDOS spectra simulations associated with the dispersion simulations in Fig.~\ref{fig:Dispersion}. These simulations were performed at 295 and 1200 K using the generalized gradient approximation (GGA + \textit{U}) density functional theory (DFT) and a Hubbard \textit{U} correction described in Ref.~\citen{pang2013}. The effect of finite temperature on the phonon structure was taken into account via the quasi-harmonic approximation as described in Ref.~\citen{pang2013}. In this approach, the equilibrium phonon structure at a given temperature is determined by free energy minimization of both the potential energy and the temperature-dependent energy of the phonon subsystem. Phonon energies are volume-dependent, hence, inclusion of the phonon energy term implicitly accounts for the anharmonic thermal expansion effects on the phonon structure.~\cite{baroni2010} UO$_\textnormal{2}$ is paramagnetic with a fluorite cubic crystal structure above 30 K. In our simulations at 295 and 1200 K, the cubic crystal symmetry rather than the distorted ground state structure~\cite{dorado2010} with the spin arrangement approximated to be the low temperature antiferromagnetic 1-\textbf{k} structure was applied. The first-principles calculations were performed using the VASP simulation package~\cite{kresse1993,*kresse1994,*kresse1996a,*kresse1996b}  and the phonon calculations were performed using PhonTS software package.~\cite{phonTS,*chernatynskiy2010} The TO-LO splitting at the zone center $\Gamma$ in UO$_\textnormal{2}$ was accounted for using the approach proposed by Wang \textit{et al.}~\cite{wang2010} The effective Born charges of 5.31\textit{e} for U$^\textnormal{4+}$ and -2.655\textit{e} for O$^\textnormal{2-}$, and dielectric constant of 5.2 needed in the simulations of the TO-LO splitting effect were determined from a separate linear response  calculation. We note that the PDOS calculations as a function of energy in Fig.~\ref{fig:PDOS_ExptDFT} were performed with $25\times25\times25$ $k$-point resolution compared with the $13\times13\times13$ resolution phonon dispersion simulations reported in Ref.~\citen{pang2013}, as a function of both wave vector and energy. 

\begin{figure}
\includegraphics[width=3.3in]{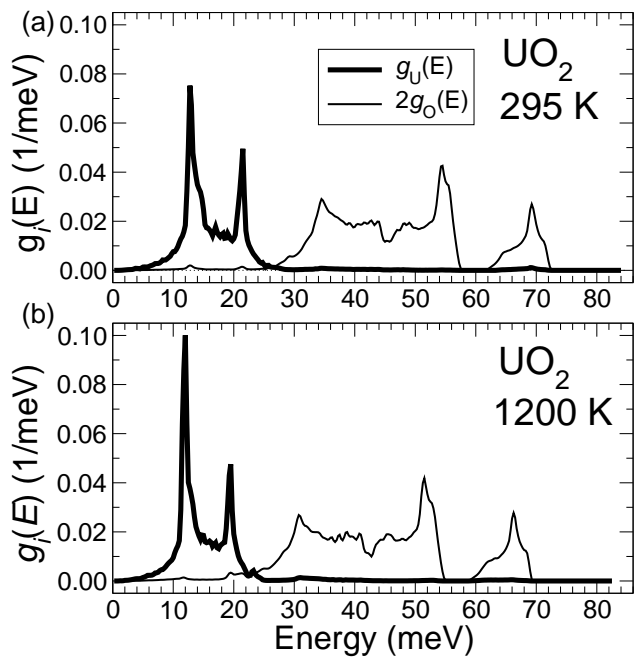}
\caption{\label{fig:PDOS_PartialDFT} Partial contribution to the UO$_\textnormal{2}$ PDOS by the one uranium (dashed line) and the two oxygen (solid line) atoms at (a) 295 K and (b) 1200 K.}
\end{figure}

In order to compare the simulated PDOS spectra with the measured PDOS spectra, the PDOS simulations in Fig.~\ref{fig:PDOS_ExptDFT} have been ``neutron-weighted" according to Eq. (\ref{eq:PDOS_NW}) using the simulated partial density of states $g_{U}(E)$  and $g_{O}(E)$ shown in Fig.~\ref{fig:PDOS_PartialDFT}. The line plots in Fig.~\ref{fig:PDOS_PartialDFT} indicate that uranium vibrations associated with acoustic phonons dominate the spectra for energies below 25 meV at 295 K and below 22 meV at 1200 K; while oxygen vibrations associated with optical phonons dominate (95 \textendash\  99 \%) the PDOS spectra for energies from 25 up to 70 meV at 295 K and from 22 to 65 meV at 1200 K. This almost complete separation of the U and O spectra in the simulations is in contrast to the experimentally observed overlapping of the LO1 and TO1 modes that have energies as low as 20 meV with the LA mode that has energies up to 25 meV at the X point along the $[001]$ direction in Fig.~\ref{fig:Dispersion}. The presence of a $\sim$ 5 meV gap in the simulated (quasi-harmonic) PDOS spectra at $\sim$ 60 meV is also at variance with the experimental observation of anharmonic linewidth closure of the gap to a spectral dip. The positions of the five observable zone boundary phonon peaks in the simulated PDOS spectra (dashed) in Fig.~\ref{fig:PDOS_ExptDFT}(a) for 295 K agree within $\sim$ 1 \textendash\  3 meV with the measured PDOS peak positions. However, for 1200 K in Fig.~\ref{fig:PDOS_ExptDFT}(b), larger discrepancies ($\sim$ 4 \textendash\  5 meV) exist between the positions of the measured and simulated transverse and longitudinal optical zone boundary peaks at 56 and 72 meV.

\subsection{Anharmonic Phonon Linewidth Broadening}

The sharpness of the zone boundary phonon peaks in the simulated PDOS spectra in Fig.~\ref{fig:PDOS_ExptDFT} is of course a result of the absence in the quasi-harmonic approximation PDOS simulations of both anharmonic phonon linewidth and spectrometer instrumental broadening. To put the simulations on the same footing as the measurements for quantitative comparison, the ARCS spectrometer resolution and the previously measured phonon linewidths~\cite{pang2013} for UO$_\textnormal{2}$ have been convoluted with the PDOS simulations as described in Appendix~\ref{sect:app}. 

\begin{figure}
\includegraphics[width=3.3in]{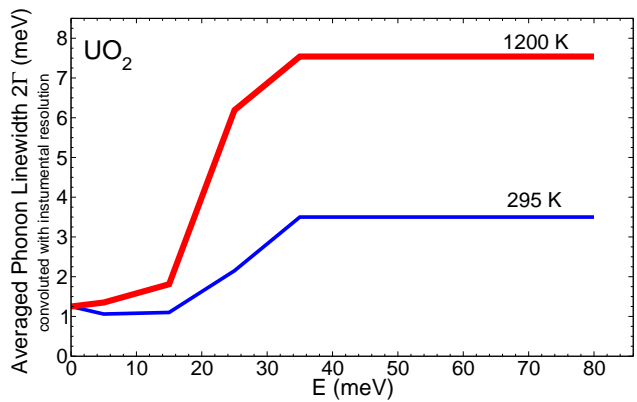}
\caption{\label{fig:FWHM} Average phonon linewidth measurements of UO$_\textnormal{2}$ as a function of energy from single crystal experiment~\cite{pang2013}, convoluted with the spectrometer ARCS instrumental resolution (see Fig.~\ref{fig:FWHM_Appen}c in Appendix~\ref{sect:app}) at 295 K (thin blue line) and 1200 K (thick red line).}
\end{figure}

Appendix~\ref{sect:app} contains plots of the measured phonon linewidths for UO$_\textnormal{2}$~\cite{pang2013} along with a tabular numerical listing of the linewidths as a function of phonon branch and wave vector, \textbf{\textit{q}}, for both 295 and 1200 K. Appendix~\ref{sect:app} also contains a description of the procedure used for energy averaging over the large (non-statistical) variances in the measured phonon linewidths. The resulting anharmonicity and instrumental broadening convoluted linewidths as a function of energy are shown in Fig.~\ref{fig:FWHM} for both 295 and 1200 K. After convoluting the (quasi-harmonic) \textit{ab initio} simulated PDOS spectra with the respective linewidth curves in Fig.~\ref{fig:FWHM}, we obtain the thick solid line spectra in Fig.~\ref{fig:PDOS_ExptDFT}, making possible a direct, quantitative comparison with the measured PDOS. For phonon energies below 25 meV, the convoluted PDOS simulations in Fig.~\ref{fig:PDOS_ExptDFT} show good agreement with the measured peak positions for the TA and LA phonons at both 295 and 1200 K. However, the spectral weights of the simulations in this energy range are markedly lower than the spectral weight in the measurements. 
For energies between 25 and 60 meV, there are small variations between the positions of the simulated and measured optical-phonon peak positions for 295 K and somewhat larger discrepancies in the spectrum for 1200 K. More noticeable is that the spectral weight of the simulated PDOS spectra in the 30 \textendash\ 60 meV range is significantly higher than the PDOS measurements for both 295 and 1200 K. We note, however, that the convolution of phonon linewidths into the simulated spectra reduces the gap at 60 meV in the simulated PDOS spectra to a dip as observed in the measured spectra. Since the area under PDOS spectra are normalized to unity, the enhancement between 30 and 60 meV is offset by the deficiencies in the spectral weights below 25 meV and above 60 meV for both the high and low temperatures. The implications of the differences between the measured and simulated PDOS spectra will be discussed below.

\section{DISCUSSION}

\subsection{Experimental PDOS measurements}
There are two aspects to note regarding the temperature dependence of the PDOS spectra shown in Fig.~\ref{fig:PDOS}. The first is that the phonon energy softening observed for 1200 K compared to 295 K does not scale as predicted within the quasi-harmonic model for phonon energies,~\cite{delaire2009,fultz2010} that is the softening does not scale with energy. For instance, the measured energy shift for the acoustic TA peak is about $\sim$ 1 meV (i.e. from 12 to 11 meV) while the corresponding energy shifts for the $\sim$ 5 times higher energy optical TO2 and LO2 phonons are also only about 1 meV, from 56 to 55 meV and 72 to 71 meV, respectively. The observation of such departures from the quasi-harmonic model question whether applying the quasi-harmonic approximations in the \textit{ab initio} simulations for UO$_\textnormal{2}$ would be able to capture the observed phonon energy softening. 

The second point is the observation of strong anharmonic broadening of the zone boundary peak at $\sim$ 33 meV in going from 295 to 1200 K; the other zone boundary peaks are impacted to a much smaller extent. The relatively broad linewidth for the TO1 zone boundary phonons at  $\sim$ 33 meV is apparently due to dispersion as indicated by the width of the TO1 phonon peak in the shell-model fitted PDOS spectrum of Dolling \textit{et al.}~\cite{dolling1965} And the large linewidth increases with temperature are in accord with the nearly a factor of 4 increase in linewidth observed for these phonons near the zone boundary X and K points ~\cite{pang2013} for UO$_\textnormal{2}$ (see TABLE~\ref{tb:EFWHM_TO1LO1}).

\begin{figure}
\includegraphics[width=3.3in]{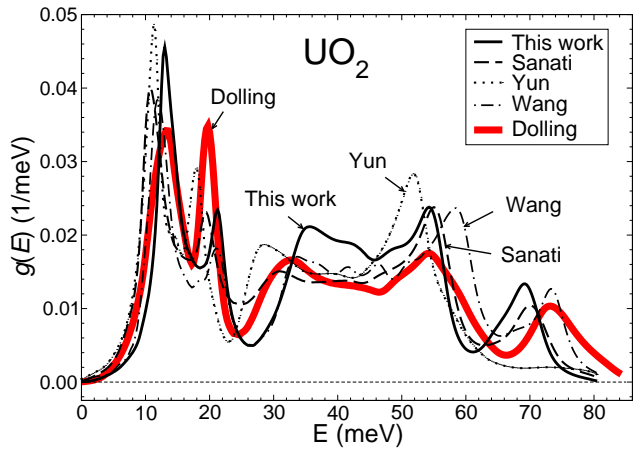}
\caption{\label{fig:PDOS_Comp} Simulated PDOS of this work using the quasi-harmonic approximation at 295 K (thin solid line); Sanati \textit{et al.}~\protect\cite{sanati2011} at 0 K (dashed line); Yun \textit{et al.}~\cite{yun2012} at 0 K with lattice parameter adjusted to the 295 K value  (dotted line); and Wang \textit{et al.}~\cite{wang_b2013} (dashed-dotted line) at 0 K;  alongside calculated PDOS (thick red line) from shell model based on phonon dispersion measurements at 295 K by Dolling \textit{et al.}~\cite{dolling1965}}
\end{figure}

For the room temperature measurements, we can compare the direct time-of-flight PDOS measurements on polycrystalline UO$_\textnormal{2}$ presented here with Dolling's shell model derived PDOS spectrum,~\cite{dolling1965} which was extracted from high-symmetry direction phonon dispersion measurements on single crystal UO$_\textnormal{2}$. Because of the reliance on phonon measurements in the high-symmetry directions only and the relatively large uncertainties for the high energy LO2 phonons in Dolling's measurements,~\cite{dolling1965} the  reliability of Dolling's shell model fitted PDOS spectrum has at times been questioned.~\cite{sanati2011,wang_b2013} In order to compare the spectra, it is necessary to neutron-weight the (non-weighted) Dolling spectrum $g_{Dolling}(E)$ in figure 5 of Ref.~\citen{dolling1965}, using Eq. (\ref{eq:PDOS_NW}) and the partial spectral weight contributions of uranium and oxygen as a function of energy. Since Dolling's shell model partial contributions were not reported, using our \textit{ab initio} simulated partial contributions at 295 K plotted in Fig.~\ref{fig:PDOS_PartialDFT}(a), we obtain the neutron-weighted PDOS:

\begin{align}
  g_{Dolling}^{NW}(E)  &= g_{Dolling}(E)\bigg \{\frac{\sigma_{U}}{M_{U}}\left [\frac{g_{U}(E)}{g_{U}(E)+2g_{O}(E)}\right ]_{DFT} \nonumber \\
    &\phantom{=}\; + 2\frac{\sigma_{O}}{M_{O}} \left [\frac{g_{O}(E)}{g_{U}(E)+2g_{O}(E)}\right ]_{DFT}\bigg \} 
\label{eq:PDOS_Dolling}
\end{align}

After convolution with the 295 K linewidths in Fig.~\ref{fig:FWHM} to include the measurement resolution and the anharmonic broadening, we obtain the solid red line in Fig.~\ref{fig:PDOS_ExptDFT}(a). The agreement between Dolling's neutron-weighted PDOS and our direct PDOS measurements is remarkably good except for the lower spectral weight dips at 25 meV and 65 meV, which we argue are a result of gaps in our \textit{ab initio} simulated partial contributions that are not present in the experimental dispersion curves of UO$_\textnormal{2}$. That is, the lower than measured spectral weight from 20 \textendash\ 30 meV in the Dolling PDOS spectrum can be attributed to the lack of an overlap of the uranium and oxygen partial PDOS in the DFT simulation (as was discussed in the Results section above). The upward energy shift of the spectral weight corresponding to LO2 phonons and the $\sim$ 4 meV higher energy cutoff at 85 meV compared to our PDOS measurement are because the shell model fit of the LO2 dispersion had steeper gradients and higher zone boundary energies than the measured dispersion, as noted already in the paper of Dolling \textit{et al.}~\cite{dolling1965} The close agreement of these two PDOS spectra demonstrates the reliability of the PDOS spectra of Dolling derived from dispersion along the three symmetry directions for UO$_\textnormal{2}$. The agreement further validates our procedure of convoluting quasi-harmonic simulated PDOS spectra with the averaged phonon linewidth functions shown in Fig.~\ref{fig:FWHM}. 

\subsection{Experimental PDOS as benchmarks for first-principles simulations for UO$_\textnormal{2}$}

The good agreement between the time-of-flight PDOS measurements (which include the entire Brillioun zone) and Dolling's PDOS (extracted from phonon dispersion in the three high symmetry directions) underscores the information available through wave vector resolved dispersion measurements. While the wave vector integrated PDOS measurements contain important information on vibrational entropy and thermodynamic properties,~\cite{fultz2010} we focus here on the insight PDOS spectra provide for UO$_\textnormal{2}$ in terms of quantitative phonon energy and dispersion gradient benchmarks~\cite{thomas2010} for \textit{ab initio} simulations in this strongly correlated material. We consider first detailed comparisons between our \textit{ab initio} simulations and our direct measurements of PDOS spectra and then extend the comparisons more generally to recently reported first-principles UO$_\textnormal{2}$ simulations.~\cite{sanati2011,yun2012,wang_b2013}

Before discussing the comparisons, we comment that the use of a $2\times2\times2$ supercell in the simulations has the limitation that it effectively approximates interatomic forces to vanish for atoms more than one lattice constant apart. However, we note that there are a number of high-symmetry \textit{k}-points in our $25\times25\times25$ grid for which the phonon wavelengths fit entirely within the supercell. These positions provide test cases with no impact of the supercell size and have been found to fit smoothly with calculations for the non-high symmetry positions of the grid.
Moreover, phonon energy calculations with $2\times2\times2$ supercells for the non-strongly correlated isostructural materials CaF$_\textnormal{2}$~\cite{verstraete2003} and CeO$_\textnormal{2}$~\cite{wang_y2013} agree well with experimental data.

The first obervation from Fig.~\ref{fig:PDOS_ExptDFT}(a) is that for energies up to 30 meV, the 295 K simulation has significantly lower spectral weights than the PDOS measurements. This apparently stems from two sources, the steeper than measured gradients of the acoustic mode dispersion below $\sim$ 20 meV and the absence in the simulation of (oxygen-dominated) optical phonons below 30 meV at 295 K. The absence of optical phonons below 30 meV is in contrast to the dip of energies to as low as 20 meV in the measured dispersion in Fig.~\ref{fig:Dispersion}(a). Moreover, because PDOS spectra are normalized to unity, the lower than measured phonon density of states below 30 meV gives rise to higher than measured optical phonon spectral weight between 30 and 55 meV in the simulation. 

Consideration of the PDOS measurements and simulations for 1200 K leads to a similar conclusion. The energies predicted at about 30 meV are $\sim$ 4 meV higher than measured for both the optical TO1 and the LO1 phonons in the $[001]$ and $[011]$ directions, coupled with flatter dispersion gradients in the simulations. Together with the simulated TO2 energies at $\sim$ 50 meV being 5 meV lower than measurements, the simulated spectral weight becomes concentrated into a smaller energy width producing a larger spectral weight between 30 and 50 meV for 1200 K than the measured result. It is interesting that the simulated PDOS spectral weight for energies below 30 meV at 1200 K in Fig.~\ref{fig:PDOS_ExptDFT}(b) is in significantly better agreement with the measured PDOS spectrum than for 295 K considering the additional reliance on the quasi-harmonic approximation. This improved agreement may be fortuitous since the better agreement for 1200 K stems from the prediction of stronger than measured phonon softening for 1200 K. The larger than measured softening of the optical modes below 30 meV, particularly for the TO1 modes, puts them in better agreement with the measured dispersion curves below 30 meV for 1200 K than for 295 K in Fig.~\ref{fig:Dispersion}(a). Overall, compared to the measured total bandwidth spanning from $\sim$ 20 \textendash\ 75 meV for both temperatures, the simulated bandwidths span $\sim$ 25 \textendash\ 70 meV for 295 K and soften to $\sim$ 20 \textendash\ 65 meV for 1200 K.

We consider now the comparison of measured and simulated PDOS spectra as quantitative tests of our lattice dynamics simulations for UO$_\textnormal{2}$, which in turn provides a sensitive test of our ability to describe the strongly correlated electronic structure of UO$_\textnormal{2}$. First-principles lattice dynamics simulations of PDOS for non-strongly correlated materials typically achieve a high level of agreement with PDOS measurements for metals (Al),~\cite{tang2010b} semi-conductors (Si),~\cite{ward2010} thermoelectrics (AgPb$_{m}$SbTe$_{2+m}$),~\cite{manley2011} intermetallics (Y$_{3}$Co)~\cite{podlesnyak2013} to oxides (ZnO).~\cite{wang_z2013} Compared to these materials, the accuracy of the current PDOS simulations for UO$_\textnormal{2}$ is rather limited. Therefore, considering the strong sensitivity of phonon linewidths and thermal transport to phonon energies and dispersion it is important to benchmark \textit{ab initio} simulations against PDOS measurements detail. 

We note that it is not possible to compare the recently reported \textit{ab initio} PDOS spectra with our neutron-weighted PDOS spectra directly without access to their partial contributions analogous to those plotted in Fig.~\ref{fig:PDOS_PartialDFT} for our simulations. However, with the above demonstration of a direct correspondence of our time-of-flight PDOS measurements with Dolling's shell model derived PDOS, the comparisons can be performed using Dolling's ``un-weighted" spectra.~\cite{dolling1965} Accordingly, after convoluting both Dolling's data and the literature-reported simulated spectra by Sanati \textit{et al.},~\cite{sanati2011} Yun \textit{et al.}~\cite{yun2012} and Wang \textit{et al.},~\cite{wang_b2013} with the average anharmonic phonon linewidths in Fig.~\ref{fig:FWHM}, spectra suitable for comparison were generated. Fig.~\ref{fig:PDOS_Comp} provides a direct PDOS benchmarking of the \textit{ab initio} PDOS spectra. 

Figure~\ref{fig:PDOS_Comp} shows anharmonic linewidth convoluted PDOS results for each of these \textit{ab initio} simulations together with the linewidth convoluted PDOS spectrum of Dolling \textit{et al}.~\cite{dolling1965} The introduction of anharmonicity into the respective PDOS spectra by convolution does not add new physics to the comparison. However, it does provide a physically based smoothing of the (non-physical) sharp spectral features of the simulated and the shell-model dependent PDOS resulting from the delta-function energy widths associated with the (quasi)-harmonic nature of the calculations. 

We note first of all that the recently published simulations show discrepancies similar to those found with our GGA + \textit{U} PDOS simulations, which perhaps is not surprising since all of them used the DFT + \textit{U} approach. At the same time, due to the sensitivity of the PDOS to the details of the simulations, all the published spectra are different from each-other, as well as from our current work. For instance, Sanati \textit{et al.}~\cite{sanati2011} used a generalized gradient spin-density approximation (GGSA) at 0 K and neglected LO-TO splitting, whereas the simulations of Wang \textit{et al.}~\cite{wang_b2013} were performed using the local density approximation with Hubbard-\textit{U}  (LDA + \textit{U}) at 0 K. On the other hand, Yun \textit{et al.}~\cite{yun2012} simulated the PDOS spectrum by adapting a spin-polarized GGA approach in which the lattice parameter was adjusted to the experimentally measured value at 295 K.  In addition, the issues relating to the ground state in the DFT + \textit{U} calculations as discussed in Ref.~\citen{dorado2013} may also influence the resulting phonon spectra.  

Overall, none of the simulations reproduce the zone boundary peak energies and spectral weight features of the experimentally measured PDOS as well as those for the non-strongly correlated examples cited above.~\cite{tang2010b,manley2011,podlesnyak2013,wang_z2013} The low energy uranium dominated TA and LA peak positions deviate by only 1 \textendash 2 meV, but the higher energy oxygen vibration dominated peaks are shifted upward (or downward) in energy by 3 \textendash 5 meV relative to the measured PDOS. Considering the PDOS measurements reported in the present work and the PDOS spectrum of Dolling in Fig.~\ref{fig:PDOS_Comp} to be essentially equivalent for comparison with the literature cited simulations, we note that all of the DFT simulations predict a 2:1 ratio for the spectral heights of the TA:LA peaks at $\sim$ 12 and 21 meV, respectively, while the experimental results show a ratio of 1:1. We observe also that the spectral weights for the optical TO2 zone boundary peak ($\sim$ 56 meV) are larger than experiment for all of the simulations. This correlates with all of the simulations (except those of Wang \textit{et al.}~\cite{wang_b2013}) yielding energies softer than measured for the high-energy LO2 optical phonons,~\cite{sanati2011,yun2012,wang_b2013} thus pushing their spectral weights to lower energies than measured.~\cite{pang2013,dolling1965} We note also that none of the dispersion simulations~\cite{pang2013,sanati2011,yun2012,wang_b2013} reproduce the experimentally observed dip of the optical TO1 and LO1 phonons energies to $\sim$ 20 meV near the zone boundary point X. Since the departures of the simulated PDOS features from the measurements are likely to be sensitive to details within the individual simulations~\cite{dorado2013} as well as the differences in the DFT approximations used, we will not discuss the details of the agreement with individual simulations further. However, the issue of the relative heights of the TA and LA peaks, the lack of a TO1 and LO1 dip in energy down to $\sim$ 20 meV near the X point, and the tendency for extra weight in the 30 \textendash 55 meV region suggests a degree of commonality in the deficiencies within the underlying physics of the simulations as well. Roughly speaking, there is a $\sim$ 10 meV compression in the overall optical mode bandwiths (energy ranges) in the simulations compared to the measurements.

Dynamical mean field theory (DMFT)~\cite{yin2008,yin2011,dorado2013}, with the inclusion of fluctuations between electronic states,  has also been used in connection with UO$_\textnormal{2}$ simulations at 1000 K to address the strongly correlated 5\textit{f} electronic structure. However, the DMFT phonon simulations by Yin \textit{et al.}~\cite{yin2008} also fail to capture the experimentally observed~\cite{pang2013} dip of the optical TO1 and LO1 phonons below 30 meV near the X zone boundary point. Moreover, the slopes, and hence the group velocities predicted for the LO1 phonons were found to be 2 \textendash 3  lower than the measurements. Another approach used to address 5\textit{f} electron issues is screened hybrid functional DFT (HF-DFT).~\cite{jollet2009,kudin2002,wen2012} While phonon dispersion and lifetime calculations have not been reported using HF-DFT, the ability to predict the lattice constant for UO$_\textnormal{2}$~\cite{kudin2002,wen2012} without parameter adjustments that are typical within DFT + \textit{U} simulations is encouraging. Accordingly, it will be interesting to compare PDOS spectra predicted by HF-DFT with the measurements. 

\section{CONCLUSIONS}
The use of time-of-flight INS measurements in connection with ab initio simulations of PDOS for UO$_\textnormal{2}$ has provided direct insight into the large impact of anharmonicity-induced linewidth broadening on the vibrational spectrum at both ambient and high temperatures. The comparisons between the experimentally measured PDOS spectra and ab inito simulations reveal significant deficiencies in the results of presently available~\cite{pang2013,sanati2011,yun2012,wang_b2013} quasi-harmonic approaches, underscoring a need for improved \textit{ab initio} approaches to simulate lattice dynamics and anharmonicity in UO$_\textnormal{2}$. The discrepancies between the measured and the simulated PDOS spectra reported here reside at the level of phonon dispersion, which, as reported previously,~\cite{pang2013} in turn strongly influence ab initio simulations of phonon linewidths. Clearly, significant progress is needed to reach in UO$_\textnormal{2}$ the level of accuracy that is currently achievable for the lattice dynamics of non-strongly correlated electron systems. Directions that present themselves are density functional perturbation theory to eliminate potential interaction range issues in the supercell approach or \textit{ab initio} molecular dynamics to account for finite temperature effects. To address the physics of the highly correlated electronic structure in a more comprehensive manner, the use of other DFT methodologies such as HF-DFT can be considered. At a higher level, DMFT with advanced solvers~\cite{amadon2012} provides the ability to address the strongly correlated 5\textit{f} electrons in UO$_\textnormal{2}$. DMFT can also potentially be used as a basis in the new approaches that are currently under development for \textit{ab initio} lattice dynamics at finite temperatures. Each of these approaches will require computationally intensive extensions to presently applied methods for UO$_\textnormal{2}$. However, depending on which methods turn out to be successful, achieving accurate PDOS simulations may represent an important step toward understanding the underlying science of the strongly correlated insulator UO$_\textnormal{2}$.
 
\begin{acknowledgments}
We thank Curtis Maples and Michael Trammell of Oak Ridge National Laboratory (ORNL) for technical support on experimental preparation. Research at ORNL was sponsored by the U.S. Department of Energy (DOE), Office of Basic Energy Sciences (BES), CMSNF, an Energy Frontier Research Center (EFRC) hosted at Idaho National Laboratory. AC and SRP are subcontractors of the U.S. Government under Contract DE-AC07-05ID14517, under the EFRC (Office of Science, BES, FWP 1356). The Research at ORNL Spallation Neutron Source was sponsored by the Scientific User Facilities Division of BES. WJLB acknowledges the Canadian Institute for Advanced Research. The urania samples were prepared at Los Alamos National Laboratory (LANL) under support by the DOE, Office of Nuclear Energy, Fuel Cycle Research and Development program. LANL is operated by Los Alamos National Security, LLC, for the National Nuclear Security Adminstration of the DOE under contract DE-AC52-06NA25396.
\end{acknowledgments}


\renewcommand{\thetable}{A\arabic{table}}
\renewcommand{\theequation}{A\arabic{equation}}

\appendix\section{\label{sect:app}Energy Averaging of Anharmonic Phonon Linewidths}
\begin{figure}
\includegraphics[width=3.3in]{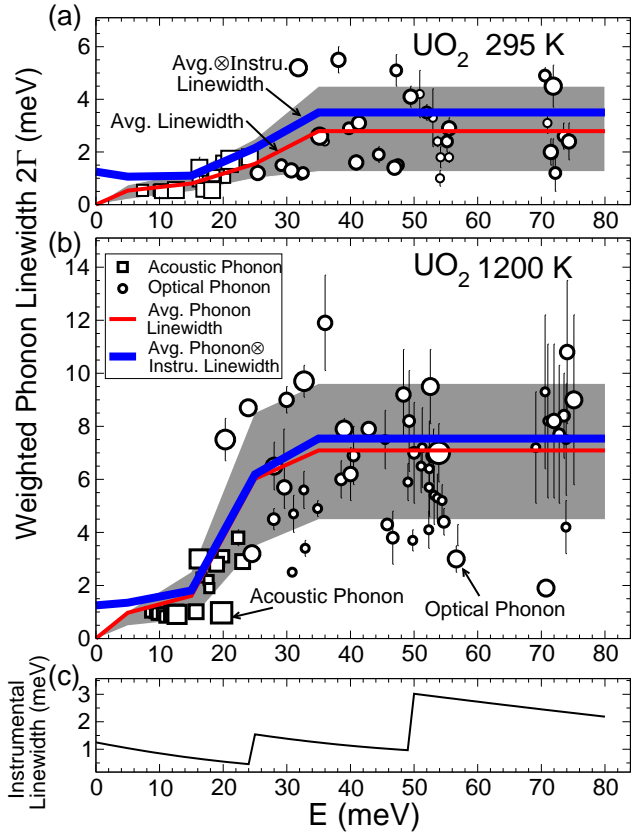}
\caption{\label{fig:FWHM_Appen} UO$_\textnormal{2}$ phonon linewidth measurements as a function of energy from single crystal experiment at (a) 295 K  and (b) 1200 K. Squares and circles represent the acoustic and optical phonons respectively. Symbol size is proportional to the weigthing of each measurement in the Brillouin zone. Thin red lines are the averaged phonon linewidths; and thick blue lines are convolution of the instrumental linewidth (shown in (c)) with the averaged phonon linewidth.}
\end{figure}

The open squares in Figs.~\ref{fig:FWHM_Appen}(a) and ~\ref{fig:FWHM_Appen}(b) represent the linewidths of the acoustic phonons, the open circles denote the linewidths of the optical phonons, and the diameters of the symbols are proportional to their volume weights given by their radial positions in phase space (i.e. proportional to $|q|^{2}$). The specification of their solid angles, $\Omega$, within the Brillouin zone, their relative weighting as a function of phonon propagation direction $[hkl]$, and their Miller index multiplicities are described in the supplementary information of Ref. ~\citen{pang2013}. Numerical values of the phonon linewidths and corresponding energies for individual phonons measured on single crystals UO$_\textnormal{2}$~\cite{pang2013} are listed in Tables~\ref{tb:EFWHM_TALA} \textendash\ ~\ref{tb:EFWHM_TO2LO2}.

\begin{figure}
\includegraphics[width=3.3in]{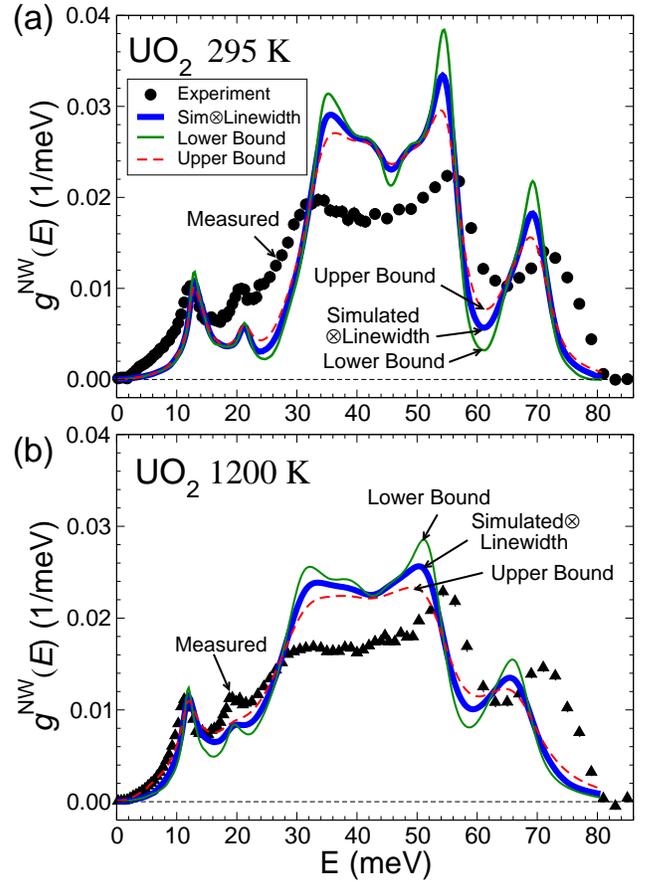}
\caption{\label{fig:PDOS_bound} Simulated PDOS using upper (red dashed curve) and lower (green solid curve) bounds of the linewidth variation with energy. Measurements are shown as symbols.}
\end{figure}

We emphasize that the large variances in the measured anharmoinc linewidths as a function of energy are not the result of measurement uncertainties; rather, the scatter is a measure of the (non-statistical) range of measured linewidths as a function of acoustic and optical phonon branches and wave vectors. Because PDOS measurements and simulations integrate over all phonon propagation directions, branches, and wave vectors, these linewidths and their variances have been accounted for using the solid line average curves in Figs.~\ref{fig:FWHM_Appen}(a) and ~\ref{fig:FWHM_Appen}(b) as determined by 10 meV binning averages of the linewidths up to 35 meV and by a flat line average between 35 and 75 meV. The energy dependence of the ``average" anharmonic linewidths for 295 and 1200 K are similar in shape but have an overall factor of $\sim$ 2.5 increase in the magnitudes of the phonon linewidths going from 295 to 1200 K. 

The relatively small linewidths ($\sim$ 0.5 \textendash\  1.0 meV) for acoustical phonons (open squares) below $\sim$ 15 meV are followed by rapidly increasing linewidths up to the LA phonon zone boundary energies of ~ 25 meV. Above 25 meV the phonon linewidths are from optical modes and tend to plateau (albeit with large variances) to about 2.8 meV (3.5 meV) for 295 K and approximately 7.1 (7.54 meV) at 1200 K. We note that the shift from open-square (acoustic) symbols below 25 meV to open-circle (optical) symbols above 25 meV in Figs.~\ref{fig:FWHM_Appen}(a) and ~\ref{fig:FWHM_Appen}(b) corresponds to the position of the change from uranium dominated partial PDOS to oxygen-dominated partial PDOS in the \textit{ab initio} simulations plotted in Fig.~\ref{fig:PDOS_PartialDFT} in the main text.
 
The combined anharmonicity and instrumental linewidths were determined by calculating for each of the linewidth points in Figs.~\ref{fig:FWHM_Appen}(a) and ~\ref{fig:FWHM_Appen}(b) the root-mean-square sum of the anharmonicity and the corresponding instrumental resolution in Fig.~\ref{fig:FWHM_Appen}(c). To demonstrate that convolution with the average linewidths in the presence of such large variances is meaningful, we plot in Figs.~\ref{fig:PDOS_bound}(a) and~\ref{fig:PDOS_bound}(b) convolutions with the upper and lower bounds of the shaded areas in Figs.~\ref{fig:FWHM_Appen}(a) and ~\ref{fig:FWHM_Appen}(b). These results verify that the uncertainties introduced by the variances are such that quantitative comparisons can be made between PDOS measurements and simulations. Such comparisons may be done not only in terms of matching PDOS peak positions with zone boundary energies, but also in terms of the phonon group velocities (i.e. phonon dispersion gradients) since the PDOS spectral weights decrease$/$increase with phonon dispersion gradients.

\bibliography{UO2_PDOS_Pang_v1}

\newpage

\begin{table*}
\parbox{.45\linewidth}{
\centering
\caption{Energies ($E$) and full-width at half-maximum (2$\mathit{\Gamma}$) of TA and LA phonons in [001], [110] and [111] directions of UO$_\textnormal{2}$ determined by single crystal measurements~\cite{pang2013} at 295 and 1200 K. Units of $E$ and 2$\mathit{\Gamma}$ are meV.}
\begin{displaymath}
    \begin{tabular}{ccccc}\hline \hline
		\multicolumn{5}{c}{TA}\\ \hline
			Wave vectors &\hspace{1.5ex}E$_{295}$\hspace{1.5ex}&\hspace{1.5ex}2$\Gamma_{295}$\hspace{1.5ex} &\hspace{1.5ex}E$_{1200}$\hspace{1.5ex}&\hspace{1.5ex}2$\Gamma_{1200}$\hspace{1.5ex} \\ \hline
(0,0,0.2)&3.82(2)&--&3.56(1)&0.88(2)\\
(0,0,0.4)&7.32(2)&0.50(1)&6.83(3)&0.91(4)\\
(0,0,0.5)&--&--&8.52(4)&0.98(4)\\
(0,0,0.6)&	10.36(2)	&	0.50(1)	&9.74(5)&0.94(8)\\
(0,0,0.7)&	--	&--&10.94(7)	&	0.80(1)\\
(0,0,0.8)&	12.57(1)	&0.55(1)&11.87(9)&1.00(1)\\
(0,0,0.9)&	--&--&12.43(9)&1.00(2)\\
(0,0,1.0)&	13.52(3)&--&12.75(9)	&0.90(2)\\
\vspace{1ex}
(0.2,0.2,0)&5.67(2)&--&5.17(2)&1.02(2)\\
(0.4,0.4,0)&11.36(1)&0.5(1)&10.64(6)&0.93(9)\\
(0.5,0.5,0)&14.36(2)&--&--&--\\
(0.6,0.6,0)&16.87(2)&0.6(2)&15.8(1)&1.0(2)\\
(0.8,0.8,0)&18.26(7)&0.5(1)&19.8(1)&1.0(2)\\
(1.0,1.0,0)&25.91(9)&--&24.2(3)&--\\
\vspace{1ex}
(0.1,0.1,0.1)&--&--&4.21(2)&0.93(3)\\
(0.2,0.2,0.2)&7.63(1)&0.5(1)&7.06(2)&0.99(5)\\
(0.3,0.3,0.3)&9.95(1)&0.5(1)&9.28(4)&0.96(5)\\
(0.4,0.4,0.4)&--&--&10.55(3)&0.94(5)\\
(0.5,0.5,0.5)&--&--&10.82(5)&--\\
\hline
\multicolumn{5}{c}{LA}\\ \hline
\textit{q} & E$_{295}$&2$\Gamma_{295}$ &E$_{1200}$&2$\Gamma_{1200}$\\\hline 
(0,0,0.2)&8.57(7)&--&7.74(4)&1.2(2)\\
(0,0,0.3)&12.88(4)&0.8(2)&12.01(3)&1.1(1)\\
(0,0,0.4)&16.07(5)&0.9(2)&14.97(3)&1.5(1)\\
(0,0,0.5)&--&--&17.70(4)&2.2(2)\\
(0,0,0.6)&21.21(5)&1.5(2)&19.97(5)&3.1(2)\\
(0,0,0.7)&--&--&22.35(8)&3.8(3)\\
(0,0,0.8)&24.19(8)&1.8(3)&23.06(5)&2.9(2)\\
(0,0,1.0)&25.9(1)&--&24.80(7)&--\\
\vspace{1ex}
(0.2,0.2,0)&11.56(5)&0.8(2)&11.06(4)&1.7(1)\\
(0.3,0.3,0)&15.58(3)&1.0(1)&14.88(3)&1.6(1)\\
(0.4,0.4,0)&18.54(5)&1.1(2)&17.82(4)&1.9(2)\\
(0.6,0.6,0)&19.98(7)&1.6(3)&18.96(5)&2.8(2)\\
(0.75,0.75,0)&--&--&16.22(5)&3.0(3)\\
(0.8,0.8,0)&16.25(5)&1.4(3)&15.30(5)&--\\
\vspace{1ex}
(0.15,0.15,0.15)&--&--&9.45(3)&1.39(9)\\
(0.2,0.2,0.2)&--&--&12.29(3)&1.9(1)\\
(0.3,0.3,0.3)&17.66(9)&0.9(1)&16.59(4)&2.3(2)\\
(0.4,0.4,0.4)&20.37(3)&1.0(1)&19.17(5)&3.0(2)\\
(0.5,0.5,0.5)&21.13(4)&1.7(3)&19.83(6)&--\\
\hline
\\
\\
\\
\\
\\
    \end{tabular}
\end{displaymath}
\label{tb:EFWHM_TALA}
}
\hfill
\parbox{.45\linewidth}{
\centering
\caption{Energies ($E$) and full-width at half-maximum (2$\mathit{\Gamma}$) of TO1 and LO1 phonons in [001], [110] and [111] directions of UO$_\textnormal{2}$ determined by single crystal measurements~\cite{pang2013} at 295 and 1200 K. Units of $E$ and 2$\mathit{\Gamma}$ are meV.}
\begin{displaymath}
    \begin{tabular}{ccccc}\hline \hline
		\multicolumn{5}{c}{TO1}\\ \hline
			Wave vectors &\hspace{1.5ex}E$_{295}$\hspace{1.5ex}&\hspace{1.5ex}2$\Gamma_{295}$\hspace{1.5ex} &\hspace{1.5ex}E$_{1200}$\hspace{1.5ex}&\hspace{1.5ex}2$\Gamma_{1200}$\hspace{1.5ex} \\ \hline
(0,0,0)&33.93(4)&--&--&--\\
(0,0,0.2)&33.78(5)&--&32.68(5)&5.6(7)\\
(0,0,0.4)&32.54(5)&1.2(2)&31.06(4)&4.7(7)\\
(0,0,0.6)&30.70(5)&1.3(2)&29.61(5)&5.7(8)\\
(0,0,0.8)&29.19(7)&1.6(3)&28.00(4)&6.5(9)\\
(0,0,1.0)&28.49(9)&--&--&--\\
\vspace{1ex}
(0.0,0.0,0)&33.5(1)&--&32.3(4)&--\\
(0.2,0.2,0)&32.24(4)&1.1(2)&30.8(1)&2.5(2)\\
(0.4,0.4,0)&29.15(6)&1.5(2)&27.9(3)&4.5(4)\\
(0.6,0.6,0)&25.43(7)&1.2(1)&24.5(2)&3.2(2)\\
(0.8,0.8,0)&22.03(5)&2.5(3)&20.84(6)&4.4(2)\\
\vspace{1ex}
(0.0,0.0,0.0)&33.5(1)&--&32.3(9)&--\\
(0.1,0.1,0.1)&--&--&32.9(1)&3.4(3)\\
(0.2,0.2,0.2)&35.98(8)&2.4(3)&34.84(7)&4.9(3)\\
(0.3,0.3,0.3)&39.75(7)&2.9(4)&38.5(2)&6.0(7)\\
(0.4,0.4,0.4)&41.3(5)&3.1(4)&40.0(3)&6(1)\\
\hline
\multicolumn{5}{c}{LO1}\\ \hline
Wave vector & E$_{295}$&2$\Gamma_{295}$ &E$_{1200}$&2$\Gamma_{1200}$\\\hline 
(0,0,0)&54.2(3)&--&--&--\\
(0,0,0.1)&53.0(2)&3.3(8)&51.1(4)&6(1)\\
(0,0,0.2)&50.9(2)&4.2(9)&49.0(2)&5.9(7)\\
(0,0,0.4)&47.2(2)&5.1(6)&45.5(2)&7(1)\\
(0,0,0.5)&--&--&40.5(3)&7(1)\\
(0,0,0.6)&38.1(1)&5.5(5)&36.0(3)&12(2)\\
(0,0,0.7)&31.8(2)&5.2(3)&30.0(1)&9.0(5)\\
(0,0,0.8)&--&--&23.9(1)&8.7(3)\\
(0,0,0.9)&--&--&20.3(1)&7.5(8)\\
(0,0,1.0)&20.80(6)&--&18.9(1)&--\\
\vspace{1ex}
(0.0,0.0,0)&56.1(3)&--&--&--\\
(0.1,0.1,0)&--&--&54.0(3)&--\\
(0.2,0.2,0)&54.6(1)&--&--&--\\
(0.3,0.3,0)&52.1(1)&--&49.8(2)&3.7(4)\\
(0.4,0.4,0)&47.47(9)&1.5(2)&45.8(2)&4.3(2)\\
(0.5,0.5,0)&44.45(8)&1.9(3)&42.9(2)&7.9(2)\\
(0.6,0.6,0)&40.91(7)&1.6(3)&39.0(1)&7.9(4)\\
(0.7,0.7,0)&35.15(6)&2.6(3)&32.7(2)&9.7(6)\\
(0.8,0.8,0)&31.4(1)&4.0(6)&29.5(2)&7.3(6)\\
(1.0,1.0,0)&27.9(2)&--&26.3(3)&--\\
\vspace{1ex}
(0.0,0.0,0.0)&56.70(4)&--&--&--\\
(0.1,0.1,0.1)&--&--&52(1)&6(1)\\
(0.2,0.2,0.2)&53.69(1)&2.4(2)&51(1)&7(1)\\
(0.3,0.3,0.3)&51.97(1)&3.5(3)&50.0(9)&7(2)\\
(0.35,0.35,0.35)&--&--&49(2)&8(2)\\
(0.4,0.4,0.4)&49.47(1)&4.1(4)&48.3(7)&9(2)\\
(0.5,0.5,0.5)&47.02(5)&--&47.1(9)&--\\
\hline
    \end{tabular}
\end{displaymath}
\label{tb:EFWHM_TO1LO1}
}
\end{table*}

\begin{table*}
\parbox{.45\linewidth}
{
\caption{Energies ($E$) and full-width at half-maximum (2$\mathit{\Gamma}$) of TO2 and LO2 phonons in [001], [110] and [111] directions of UO$_\textnormal{2}$ determined by single crystal measurements~\cite{pang2013} at 295 and 1200 K. Units of $E$ and 2$\mathit{\Gamma}$ are meV.}
\begin{displaymath}
    \begin{tabular}{ccccc}\hline \hline
		\multicolumn{5}{c}{TO2}\\ \hline
			Wave vectors &\hspace{1.5ex}E$_{295}$\hspace{1.5ex}&\hspace{1.5ex}2$\Gamma_{295}$\hspace{1.5ex} &\hspace{1.5ex}E$_{1200}$\hspace{1.5ex}&\hspace{1.5ex}2$\Gamma_{1200}$\hspace{1.5ex} \\ \hline
(0,0,0.2)&54.2(2)&1.8(4)&52.4(4)&5.7(9)\\
(0,0,0.4)&55.2(2)&2.4(4)&53.7(4)&5.3(9)\\
(0,0,0.6)&55.5(2)&2.9(4)&53.1(4)&7(1)\\
(0,0,0.8)&55.4(1)&2.2(5)&52.5(5)&10(1)\\
(0,0,1.0)&--&--&53.9(4)&7(1)\\
\vspace{1ex}
(0.0,0.0,0)&55.7(2)&--&--&--\\
(0.2,0.2,0)&55.5(2)&1.8(2)&54.4(5)&5.2(6)\\
(0.4,0.4,0)&56.6(2)&--&54.7(2)&4.4(5)\\
(0.6,0.6,0)&58.5(2)&--&56.6(4)&3.0(5)\\
(0.8,0.8,0)&59.0(3)&--&56.9(6)&3.9(4)\\
\vspace{1ex}
(0.0,0.0,0.0)&56.2(6)&--&--&--\\
(0.1,0.1,0.1)&--&--&52.3(3)&4.1(7)\\
(0.2,0.2,0.2)&54.1(2)&1.0(2)&53.2(2)&5(1)\\
(0.35,0.35,0.35)&--&--&46.7(6)&3.8(7)\\
(0.4,0.4,0.4)&46.92(4)&1.4(1)&--&--\\
(0.5,0.5,0.5)&46.0(2)&--&--&--\\
\hline
\multicolumn{5}{c}{LO2}\\ \hline
Wave vector & E$_{295}$&2$\Gamma_{295}$ &E$_{1200}$&2$\Gamma_{1200}$\\\hline 
(0,0,0.0)&71.1(6)&--&71(1)&--\\
(0,0,0.2)&71.9(4)&1.9(6)&70.6(6)&9(2)\\
(0,0,0.4)&73.5(2)&2.6(5)&73.9(4)&8(2)\\
(0,0,0.6)&74.3(3)&2.4(7)&74.1(5)&11(2)\\
(0,0,0.8)&75.7(4)&--&75.1(9)&9(3)\\
(0,0,1.0)&76.2(6)&--&77(1)&--\\
\vspace{1ex}
(0.0,0.0,0)&70.5(4)&--&71(1)&--\\
(0.2,0.2,0)&74.5(7)&--&73.9(7)&4(1)\\
(0.4,0.4,0)&72.2(4)&1.2(4)&73.6(6)&8(1)\\
(0.6,0.6,0)&71.5(1)&2.0(5)&70.7(2)&1.9(3)\\
(0.8,0.8,0)&--&--&73.1(3)&--\\
(1.0,1.0,0)&76(1)&--&75.1(9)&--\\
\vspace{1ex}
(0,0,0)&69.7(4)&--&69(1)&--\\
(0.1,0.1,0.1)&71.0(2)&3.1(4)&69.1(8)&7.2(2)\\
(0.2,0.2,0.2)&--&--&71.0(9)&8.2(2)\\
(0.3,0.3,0.3)&70.6(2)&4.9(3)&72.8(9)&7.7(2)\\
(0.4,0.4,0.4)&71.9(2)&4.5(7)&72(1)&8.2(3)\\
(0.5,0.5,0.5)&72.7(4)&--&73(1)&--\\
\hline

    \end{tabular}
\end{displaymath}
\label{tb:EFWHM_TO2LO2}
}
\end{table*}


\end{document}